\documentclass[copyright,creativecommons,noncommercial,sharealike]{eptcs}
\providecommand{\event}{GandALF 2017} %

\makeatletter{}%

\usepackage[utf8]{inputenc}

\usepackage{amssymb}
\usepackage{amsfonts}
\usepackage[tbtags]{amsmath}
\usepackage{amsthm}
\usepackage{url}
\usepackage{underscore}

\makeatletter

\thm@headfont{%
  \textcolor{red}{$\blacktriangleright$}\nobreakspace\sffamily\bfseries}
\def\th@remark{%
  \thm@headfont{\bfseries%
    \textcolor{darkgray}{$\blacktriangleright$}\nobreakspace\sffamily}%
  \normalfont %
}
\def\@endtheorem{\endtrivlist}%

\newenvironment{sketch}[1][Sketch of proof]{\par
  \pushQED{\qed}%
  \normalfont \topsep6\p@\@plus6\p@\relax
  \trivlist
  \item[\hskip\labelsep
        \color{darkgray}\sffamily\bfseries
    #1\@addpunct{.}]\ignorespaces
}{%
  \popQED\endtrivlist%
}
\theoremstyle{plain}
\newtheorem{theorem}{Theorem}
\newtheorem{lemma}[theorem]{Lemma}

\theoremstyle{definition}

\theoremstyle{remark}

\ifx\numberwithinsect\relax
  \@addtoreset{theorem}{section}
  \edef\thetheorem{\expandafter\noexpand\thesection\@thmcountersep\@thmcounter{theorem}}
\fi

\makeatother

\usepackage{subcaption}
\usepackage[dvipsnames, usenames]{xcolor}
\usepackage{tikz}
\usetikzlibrary{calc, backgrounds, decorations.pathmorphing, fit}
\usepackage{lscape}
\usepackage{stmaryrd}   %

\usepackage[normalem]{ulem}

\usepackage{algorithm}
\usepackage[noend]{algpseudocode}
\algrenewcommand\algorithmicrequire{\textbf{Input:}}
\algrenewcommand{\algorithmicensure}{\textbf{Output:}}
\algnewcommand\algorithmicswitch{\textbf{switch}}
\algnewcommand\algorithmiccase{\textbf{case}}
\algnewcommand\algorithmicassert{\texttt{assert}}
\algnewcommand\Assert[1]{\State \algorithmicassert(#1)}%
\algdef{SE}[SWITCH]{Switch}{EndSwitch}[1]{\algorithmicswitch\ #1\ :}{\algorithmicend\ \algorithmicswitch}%
\algdef{SE}[CASE]{Case}{EndCase}[1]{\algorithmiccase\ #1 :}{\algorithmicend\ \algorithmiccase}%
\algtext*{EndSwitch}%
\algtext*{EndCase}%

\usepackage{numprint}
\npdecimalsign{.}
\npthousandsep{,}

\usepackage[draft = false]{microtype}
 
\makeatletter{}\newcommand{\Nat}{\mathbb{N}}
\newcommand{\Rational}{\mathbb{Q}}
\newcommand{\Real}{\mathbb{R}}

\newcommand{\tudparagraph}[2]{%
\vspace*{#1}

\noindent
\textbf{#2}
}

\newcommand{\cL}{\mathcal{L}}
\newcommand{\cM}{\mathcal{M}}
\newcommand{\cO}{\mathcal{O}}

\newcommand{\fC}{\mathfrak{C}}
\newcommand{\fM}{\mathfrak{M}}
\newcommand{\fN}{\mathfrak{N}}

\newcommand{\eqdef}{\mathrel{\stackrel{\text{\tiny def}}{=}}}

\newcommand{\Expect}{\mathrm{E}}
\newcommand{\Prob}{\mathrm{Pr}}
\renewcommand{\Pr}{\Prob}

\newcommand{\sinit}{s_{\textit{\tiny init}}}
\newcommand{\bfP}{\mathbf{P}}
\newcommand{\AP}{\mathrm{AP}}

\newcommand{\Post}{\mathit{Post}}
\newcommand{\Constr}{\mathit{Constr}}

\DeclareMathOperator{\degree}{deg}

\DeclareMathOperator{\Paths}{Paths}

\newcommand{\wgt}{\mathit{wgt}}

\newcommand{\Goal}{\mathit{Goal}}

\newcommand{\goal}{\mathit{goal}}

\newcommand{\true}{\texttt{true}}
\newcommand{\false}{\texttt{false}}

\newcommand{\POp}[1][\bowtie c]{\operatorname{\mathbb{P}}_{#1}}
\newcommand{\EOp}[1][\bowtie r]{\operatorname{\mathbb{E}}_{#1}}
\newcommand{\COp}[1]{\operatorname{\mathbb{C}}_{#1}}

\newcommand{\PrOp}{\mathbb{P}}

\newcommand{\CompOp}{\mathbb{C}}

\renewcommand{\Until}{\mathbin{\mathsf{U}}}
\newcommand{\Release}{\mathbin{\mathsf{R}}}
\DeclareMathOperator{\neXt}{\bigcirc}
\DeclareMathOperator{\Eventually}{\Diamond}

\DeclareMathOperator{\accdiaplus}{\tikz[baseline = -0.65ex]{\node [inner sep = 0pt] {$\Diamond$}; \draw (0,-0.9ex) -- (0,0.9ex) (-0.6ex,0) -- (0.6ex,0);}}
\DeclareMathOperator{\MeanPayoff}{mp}

\newcommand{\Sat}{\mathrm{Sat}}
\newcommand{\tSat}{\texttt{Sat}}

\newcommand{\psec}[1]{\nprounddigits{2}\npfourdigitnosep\numprint[s]{#1}}
\newcommand{\timeout}{time-out}

\newcommand{\selim}{\emph{state-elim}}
\newcommand{\geff}{\emph{GE-ff}}
\newcommand{\geffred}{\emph{\mbox{red(GE-ff)}}}
\newcommand{\gefac}{\emph{GE-fac}}
\newcommand{\eigen}{\emph{eigen}}

\newcommand{\citeAppendix}[1]{}
\newcommand{\citeExtended}[1]{}

\renewcommand{\citeExtended}[1]{#1}

\makeatletter{}
\title{Parametric Markov Chains:
       PCTL Complexity and Fraction-free Gaussian Elimination%
\footnote{%
The authors are supported by the DFG through
        the Collaborative Research Center SFB 912 -- HAEC,
        the Excellence Initiative by the German Federal and State Governments (cluster of excellence cfaed),
        the Research Training Group QuantLA (GRK 1763)
        and the DFG-projects BA-1679/11-1 and BA-1679/12-1.
}
}

\author{Lisa Hutschenreiter
  \and Christel Baier
  \and Joachim Klein
\institute{Technische Universit\"at Dresden, Dresden, Germany}
\email{\{Lisa.Hutschenreiter,Christel.Baier,Joachim.Klein\}@tu-dresden.de}
}

\def\titlerunning{Parametric Markov Chains: PCTL Complexity and Fraction-free Gaussian Elimination}
\def\authorrunning{L. Hutschenreiter, C. Baier \& J. Klein}

\begin{document}

\maketitle

\makeatletter{}
\begin{abstract}
Parametric Markov chains have been introduced as a model for families of
stochastic systems that rely on the same graph structure, but differ in the
concrete transition probabilities. The latter are specified by polynomial
constraints for the parameters. Among the tasks typically addressed in the
analysis of parametric Markov chains are (1) the computation of closed-form
solutions for reachabilty probabilities and other quantitative measures
and (2) finding symbolic representations of the set of parameter valuations
for which a given temporal logical formula holds as well as (3) the decision
variant of (2) that asks whether there exists a parameter valuation where a
temporal logical formula holds. Our contribution to (1) is to show that
existing implementations for computing rational functions for reachability
probabilities or expected costs in parametric Markov chains can be improved
by using fraction-free Gaussian elimination, a long-known technique for linear
equation systems with parametric coefficients. Our contribution to (2) and (3)
is a complexity-theoretic discussion of the model checking problem for
parametric Markov chains and probabilistic computation tree logic (PCTL)
formulas. We present an exponential-time algorithm for (2) and a PSPACE upper
bound for (3). Moreover, we identify fragments of PCTL and subclasses of
parametric Markov chains where (1) and (3) are solvable in polynomial time
and establish NP-hardness for other PCTL fragments.
\end{abstract}

\makeatletter{}\section{Introduction}

Finite-state Markovian models are widely used as an operational model for
the quantitative analysis of systems with probabilistic behaviour.
In many cases, only estimates of the transition probabilities are
available. This, for instance, applies to fault-tolerant systems
where the transition probabilities are derived from error models 
obtained using statistical methods. Other examples are systems operating with
resource-management protocols that depend on stochastic assumptions 
on the future workload, or cyber-physical systems where the interaction
with its environment is represented stochastically.
Furthermore, often the transition probabilities of Markovian models
depend on configurable system parameters that can be adjusted at design-time.
The task of the designer is to find a parameter setting 
that is optimal with respect to a given objective.
This motivated the investigation of 
\emph{interval Markov chains} (IMCs) \cite{JonLar91} 
specifying intervals for the transition probabilities 
(rather than concrete values).
More general is the model of \emph{parametric Markov chains} (pMCs), 
which has been
introduced independently by Daws \cite{Daws05} and  
Lanotte et al.~\cite{LanMagSchTroina07},
where the transition probabilities are given by
polynomials
with rational coefficients 
over a fixed set of real-valued parameters $x_1,\ldots,x_k$.
These concepts can be further generalized to accommodate
rational functions,
i.\,e., quotients of polynomials, as transition
probabilities~(see, e.\,g.,~\cite{HHZ-STTT11}).

It is well-known that the probabilities $p_s$ for
reachability conditions $\Diamond \Goal$ in parametric Markov chains
with a finite state space $S$ 
can be characterized as the unique solution of a linear equation
system $A \cdot p = b$
where $p=(p_s)_{s\in S}$ is the solution vector, 
and
$A = A(x_1,\ldots,x_k)$ is a matrix where the coefficients
are rational functions.
Likewise,
$b= b(x_1,\ldots,x_k)$ is a vector whose coefficients are
rational functions.
Note that it is no limitation to assume that the entries in $A$
and $b$ are polynomials, as rational function entries can be converted
to a common denominator, which can then be removed.
Now, $A \cdot p = b$ can be viewed as a linear equation system
over the field $\Rational(x_1,\ldots,x_k)$ 
of rational functions with rational coefficients.
As a consequence, the probabilities for reachability conditions are 
rational functions.
This has been observed independently by Daws~\cite{Daws05} and 
Lanotte et al.~\cite{LanMagSchTroina07} 
for pMCs.
Daws~\cite{Daws05} describes a computation
scheme that relies on a state-elimination algorithm inspired
by the state-elimination algorithm for computing regular expressions
for nondeterministic finite automata. This, however, is fairly the same
as Gaussian elimination for matrices over the field of rational 
functions. 

As observed by Hahn et al.~\cite{HHZ-STTT11},
the na\"ive implementation of Gaussian elimination for pMCs,
that treats the polynomials in $A$ and $b$ as syntactic atoms,
leads to a representation of the rational functions $p_s=p_s(x_1,\ldots,x_k)$
as the quotient of extremely (exponentially) large polynomials.
In their implementation PARAM~\cite{PARAM-HHWZ10}
(as well as in the re-implementation within the tool PRISM~\cite{KwiatkowskaNP11}),
the authors of~\cite{HHZ-STTT11} 
use computer-algebra tools
to simplify rational functions in each step of Gaussian elimination
by identifying the greatest common divisor (gcd) of the numerator and the
denominator polynomial.
Together with polynomial-time algorithms for the gcd-computation of
univariate polynomials,
this approach 
yields a polynomial-time algorithm for computing the rational functions
for reachability probabilities in pMCs 
with a single parameter.
Unfortunately, gcd-computations are known to be expensive 
for the multivariate case (i.\,e., $k \geqslant 2$) \cite{GeCzLa93}.
To mitigate the cost of the gcd-computations, the tool
Storm~\cite{DJKV-CAV17}
successfully uses techniques proposed in~\cite{JansenCVWAKB14}
such as caching and the representation of the polynomials in
partially factorized form during the elimination steps. 
However, it is possible to completely avoid gcd-computations by using
\emph{one-step fraction-free Gaussian elimination}.
Surprisingly, this has not yet been investigated 
in the context of pMCs, although
it is a well-known technique in mathematics.
According to Bareiss \cite{Bareiss72}, this variant of
Gaussian elimination probably goes back to
Camille Jordan (1838--1922), and has been rediscovered several times since.
Like standard Gaussian elimination it relies on the triangulation of the
matrix, and finally obtains the solution by back substitution. 
Applied to matrices over polynomial rings the approach
generates  matrices with polynomial coefficients (rather than
rational functions) and ensures that the degree of the polynomials 
in all intermediate matrices grows at most linearly.
This is achieved by dividing, in each elimination step, by a factor
known by construction. 
Thus, when applied to a pMC with linear expressions
for the transition probabilities,
the degree of all polynomials in the solution vector is bounded
by the number of states.
For the univariate case ($k=1$), this yields an alternative 
polynomial-time algorithm
for the computation of the rational functions for reachability probabilities.
Analogous statements hold for expectations of random variables
that are computable via linear equation systems. This applies to
expected accumulated weights until reaching a goal, and to the expected
mean payoff.

\tudparagraph{1.0ex}{Contribution.}
The purpose of the paper is to study the complexity of the
model checking problem for pMCs and
probabilistic computation tree logic (PCTL) \cite{HaJo94},
and its extensions by expectation operators for pMCs
augmented by weights for its states.
In the first part of the paper (Section \ref{sec:gauss}), 
we discuss the use of Bareiss' one-step fraction-free
Gaussian elimination for the computation of reachability probabilities.
The second part of the paper (Section \ref{sec:theory})
presents complexity-theoretic
results for the PCTL model checking problem in pMCs. 
We describe an exponential-time algorithm for computing a 
symbolic representation of all parameter valuations under which a given
PCTL formula holds, and provide a PSPACE upper bound for the
decision variants that ask whether  a given PCTL formula holds for some
or all admissible parameter valuations.
The known NP-/coNP-hardness
results for IMCs \cite{SeViAg06,ChatSenHen08} 
carry over
to the parametric case. We strengthen this result by showing that
the existential PCTL model checking problem 
remains NP-hard even for acyclic pMCs
and PCTL formulas with a single probability operator.
For the univariate case, we 
prove NP-completeness for the existential PCTL model checking problem,
and identify two fragments of PCTL where the
model checking is solvable in polynomial time.
The first fragment are Boolean combinations of
threshold constraints for reachability probabilities,
expected accumulated weights until reaching a goal, and
expected mean payoffs.
The second fragment consists of PCTL formulas in positive normal form
with lower probability thresholds interpreted over
pMCs satisfying some monotonicity
properties.
Furthermore, we observe that the model checking problem for 
PCTL with expectation operators
for reasoning about expected costs until reaching a goal
is in P for Markov chains where the weights of the states 
are given as polynomials over a single parameter, 
when restricting to Boolean combinations of the expectation
operators. 

Proofs and further details on the experiments omitted
in the main part due to space constraints can be found in the
\citeAppendix{~\hyperref[sec:appendix]{appendix} of this extended version~\cite{GandALF-extended}. }%
\citeExtended{extended version \cite{GandALF-extended}. }%

\tudparagraph{1.0ex}{Related work.}
Fraction-free Gaussian elimination
is well-known in mathematics, and has been further investigated 
in various directions for matrices over 
unique factorization domains (such as polynomial rings), see e.\,g.
\cite{McClellan73,Kannan85,Sit92,NakTurWil97}.
To the best of our knowledge, fraction-free Gaussian elimination
has not yet been studied in the context of parametric Markovian
models.

Besides the above mentioned work 
\cite{Daws05,PARAM-HHWZ10,HHZ-STTT11,JansenCVWAKB14,DJJCVBKA-CAV15}
on the computation of the rational
functions for reachability probabilities
in pMCs,
\cite{LanMagSchTroina07} identifies instances where the 
parameter synthesis problem for
pMCs with 1 or 2 parameters and  
probabilistic reachability constraints
is solvable in polynomial time.
These rely on the fact that there are closed-form representations of
the (complex) zero's for univariate polynomials up to degree 4
and rather strong syntactic characterizations of pMCs.
In Section \ref{sec:gauss} we will provide an example to illustrate
that the number of monomials in the numerators of the 
rational functions for reachability probabilities can grow exponentially
in the number of states.
We hereby reveal a flaw in \cite{LanMagSchTroina07} where
the polynomial-time computability of the rational functions
for reachability probabilities has been stated even for the multivariate
case.
\cite{FilieriGT11} considers an approach for solving the parametric linear
equation system obtained from sparse pMCs via Laplace expansion.

Model checking problems for IMCs 
and temporal logics have been studied by several authors.
Most in the spirit of our work on the complexity of the PCTL
model checking problem for pMCs is the paper~\cite{SeViAg06}
which studies the complexity of PCTL
model checking in IMCs.
Further complexity-theoretic results of the model checking problem 
for IMCs and temporal logics have been established
in~\cite{ChatSenHen08} 
for omega-PCTL 
(extending PCTL by Boolean combinations of B\"uchi and 
co-B\"uchi conditions),
and in~\cite{BLW-TACAS13} for linear temporal logic (LTL).
Our results of the second part can be seen as an extension of the work 
\cite{SeViAg06,ChatSenHen08} for the case of pMCs.
The  NP lower bound for the multivariate case and a single
threshold constraint for reachability probabilities strengthen the NP-hardness
results of \cite{SeViAg06}.

There exist several approaches to obtain regions of parameter
valuations of a pMC in which PCTL formulas are satisfied or not,
resulting in an approximative covering of the parameter space.
PARAM~\cite{HHZ-STTT11,PARAM-HHWZ10} employs a heuristic,
sampling based approach, while PROPhESY~\cite{DJJCVBKA-CAV15}
relies on SMT solving via the existential theory of the reals
to determine whether a given formula holds for all valuations
in a sub region.
For the same problem, \cite{QuatmannD0JK16} uses a parameter lifting technique
that avoids having to solve the parametric equation system by
obtaining lower and upper bounds for the values in a given region
by a reduction to non-parametric Markov decision processes.

\makeatletter{}
\section{Preliminaries}
\label{sec:prelim}

The definitions in this section require a general understanding
of Markov models, standard model checking, and temporal logics.
More details
can be found, e.\,g., in \cite{Ku95,BaKa08}.

\tudparagraph{1.0ex}{Discrete-time Markov chain.}
A \emph{(discrete-time) Markov chain} (MC) 
$\cM$ is a tuple $( S , \sinit, E, P)$ where
$S$ is a non-empty, finite set of \emph{states}
with the \emph{initial state} $\sinit\in S$,
$E \subseteq S \times S$ is a transition relation, and
$P \colon S \times S \to [0,1]$ 
is the \emph{transition probability function}
satisfying $P(s,t)=0$ if and only if $(s,t)\notin E$, and
$\sum _{t\in S } P(s,t) = 1$ for all $s\in S $ with
$\Post(s)\eqdef \{t\in S : (s,t)\in E\}$ nonempty.
We refer to $G_{\cM}=(S,E)$ as the \emph{graph} of $\cM$.
A state $s \in S$ in which $\Post(s) = \varnothing$
is called a \emph{trap (state)} of $\cM$.

An \emph{infinite path} in $\cM$ is an infinite sequence 
$s_0s_1 \ldots \in  S^{\omega}$ of states such that
$(s_i,s_{i+1})\in E$ for $i\in\Nat$.
Analogously, a \emph{finite path} in $\cM$ is a finite sequence 
$s_0s_1 \ldots s_m \in  S^{*}$ of states in $\cM$ such that
$(s_i,s_{i+1})\in E$ for $i=0,1,\ldots,m{-}1$.
A path is called \emph{maximal} if it is infinite or ends in a trap.
$\Paths(s)$ denotes the set of all maximal paths
in $\cM$ starting in $s$.
Relying on standard techniques, every MC induces a unique
probability measure $\Pr^{\cM}_s$ on the set of all paths.

\tudparagraph{1.0ex}{Parameters, polynomials, and rational functions.}
Let $x_1,\ldots,x_k$ be parameters 
that can assume any real value,
$\overline{x}=(x_1,\ldots,x_k)$.
We write $\Rational[\overline{x}]$ 
for the \emph{polynomial ring} over the rationals
with variables $x_1,\ldots,x_k$. Each $f\in \Rational[\overline{x}]$ 
can be written as a sum of monomials, i.\,e.,
$f=\sum_{(i_1,\ldots,i_k) \in I} 
  \alpha_{i_1,\ldots,i_k} \cdot 
  x_1^{i_1} \cdot x_2^{i_2} \cdot  \ldots \cdot x_k^{i_k}$
where $I$ is a finite subset of $\Nat^k$ and
$\alpha_{i_1,\ldots,i_k}\in \Rational$.
If $I$ is empty, or $\alpha_{i_1,\ldots,i_k}=0$ for all tuples $(i_1,\ldots,i_k)\in I$,
then $f$ is the \emph{null function}, generally denoted by 0.
The \emph{degree} of $f$ is 
$\degree(f) = \max \bigl\{\,  i_1 + \ldots + i_k :\allowbreak
      (i_1,\ldots,i_k)\in I, \alpha_{i_1,\ldots,i_k}\not= 0\,	\bigr\}$
where $\max ( \varnothing ) = 0$.
A \emph{linear function} is a function
$f \in \Rational{}[\overline{x}]$ with
$\deg(f) \leqslant 1$.
A \emph{rational function} is a function of the form
$f/g$ with $f,g\in \Rational[\overline{x}]$, $g \neq 0$.
The field of all rational functions
is denoted by $\Rational(\overline{x})$.
We write $\Constr[\overline{x}]$ for the set of all \emph{polynomial constraints}
of the form $f \bowtie g$ where $f,g\in \Rational[\overline{x}]$, and
$\bowtie \in \{<,\leqslant, >,\geqslant, = \}$.

\tudparagraph{1.0ex}{Parametric Markov chain.}
A \emph{(plain) parametric Markov chain} on $\overline{x}$, pMC for short,
is a tuple $\fM = ( S , \sinit, E,\bfP)$ 
where $S$, $\sinit$, and $E$ are defined as for MCs, and 
$\bfP\colon S\times S \to \Rational(\overline{x})$
is the transition probability function
with $\bfP(s,t) = 0$, i.\,e., the null function, iff $(s,t)\notin E$.
Intuitively, a pMC defines the family of Markov chains arising by plugging in 
concrete values for the parameters.
A parameter valuation $\overline{\xi} = (\xi_1,\ldots,\xi_k)\in \Real^k$
is said to be \emph{admissible} for $\fM$ if
for each state $s\in S$ we have
  $\sum_{t\in S} P_{\overline{\xi}}(s,t) =1$ if $\Post(s)$ nonempty, and
  $P_{\overline{\xi}}(s,t) >0$ iff $(s,t)\in E$,
  where $P_{\overline{\xi}}(s,t) = \bfP(s,t)(\overline{\xi})$ 
  for all $(s,t)\in S\times S$.
Let $X_{\fM}$, or briefly $X$, 
denote the set of admissible parameter valuations for $\fM$.
Given $\overline{\xi}\in X$
the Markov chain associated with $\overline{\xi}$ is
$\cM_{\overline{\xi}} = \fM(\overline{\xi}) =
    (S,\sinit,E,P_{\overline{\xi}})$.
The semantics of the pMC $\fM$ is then defined as the family of Markov chains
induced by admissible parameter valuations, i.\,e.,
  $\llbracket \fM \rrbracket =
    \bigl\{\,
      \fM(\overline{\xi}) : \overline{\xi} \in X
    \,\bigr\} $.

An \emph{augmented pMC} is a tuple $\fM = ( S, \sinit, E, \bfP, \fC )$
where $S$, $\sinit$, $E$, and $\bfP$ are defined as for plain pMCs, and
$\fC \subset \Constr[\overline{x}]$ is a
finite set of polynomial constraints.
A parameter valuation $\overline{\xi}$ is \emph{admissible}
for an augmented pMC if it is admissible for the induced
plain pMC $(S, \sinit, E, \bfP)$,
and satisfies all polynomial constraints in $\fC$.
As for plain pMC, we denote the set of admissible
parameter valuations of an augmented pMC by 
$X_{\fM}$, or briefly $X$.

A, possibly augmented, pMC $\fM$ is called \emph{linear},
or \emph{polynomial},
if all transition probability functions and
constraints are linear functions in $\overline{x}$,
or polynomials in $\overline{x}$, respectively.

\tudparagraph{1.0ex}{Interval Markov chain.}
An \emph{interval Markov chain} (IMC) \cite{SeViAg06}
can be seen as a special case of a linear augmented 
pMC with one parameter $x_{s,t}$ for each
edge $(s,t)\in E$, and linear constraints
$\alpha_{s,t} \unlhd_1 x_{s,t} \unlhd_2 \beta_{s,t}$ for each edge
with $\alpha_{s,t},\beta_{s,t}\in \Rational \cap [0,1]$ and
$\unlhd_1,\unlhd_2\in \{<,\leqslant\}$.
According to the terminology introduced in \cite{SeViAg06},
this corresponds to the semantics of IMC as an ``uncertain Markov chain''.
The alternative semantics of IMC as a Markov decision process will not be
considered in this paper.

\tudparagraph{1.0ex}{Labellings and weights.}
Each of these types of Markov chain, whether MC, plain or augmented pMC, or IMC, can be
equipped with a \emph{labelling function} $\cL \colon S \to 2^\AP$,
where $\AP$ is a finite set of \emph{atomic propositions}.
If not explicitly stated, we assume the implicit labelling of the Markov chain
defined by using the state names as atomic propositions
and assigning each name to the respective state.
Furthermore, we can extend any Markov chain with a \emph{weight
function} $\wgt\colon S\to \Rational$. The value assigned to a specific
state $s\in S$ is called the weight of $s$.
It is sometimes also referred to as the \emph{reward} of $s$.
In addition to assigning
rational values we also consider parametric weight functions
$\wgt\colon S\to \Rational( \overline{x})$.

\tudparagraph{1.0ex}{Probabilistic computation tree logic.}
We augment the standard notion of probabilistic computation tree logic 
with operators for the expected 
accumulated weight and mean payoff, and for comparison.
Let $\AP$ be a finite set of atomic propositions.
$\bowtie$ stands for $\leqslant, \geqslant, <, >$, or $=$, $c\in [0,1]$, 
$r\in\mathbb{Q}$. Then
\begin{align*}
\begin{array}{lcll}
	\Phi &::= & 
		\true \mid a \mid \Phi \wedge \Phi \mid \neg \Phi 
		\mid \POp \bigl(\varphi\bigr) \mid \EOp \bigl(\rho\bigr)
		\mid \COp{\Prob}(\varphi,\bowtie,\varphi) 
                \mid \COp{\Expect}(\rho,\bowtie,\rho)
			& \text{\footnotesize\emph{state formula}} \\[1ex]
	\varphi &::= &
		\neXt\Phi \mid \Phi\Until\Phi 
		  \hspace*{0.5cm}\text{\footnotesize\emph{path formula}} 
      \hspace*{1.5cm}
	\rho \ ::=\ 
		\accdiaplus \Phi \mid \MeanPayoff (\Phi)
			& \multicolumn{1}{r}{\hspace*{-2cm}
             \text{\footnotesize\emph{terms for random variables}}}
\end{array}
\end{align*}
where $a\in \AP$. The basic temporal modalities are $\neXt$ (\emph{next})
and $\Until$ (\emph{until}).
The usual derived temporal modalities 
$\Diamond$ (\emph{eventually}), 
$\Release$ (\emph{release}) and $\Box$ (\emph{always}) are
defined by $\Eventually \Phi \eqdef \true \Until \Phi$,
and $\POp[\bowtie c](\Phi_1 \Release \Phi_2) \eqdef 
     \POp[\overline{\bowtie} 1 {-} c] ((\neg \Phi_1) \Until (\neg \Phi_2))$,
where, e.\,g., $\overline{\leqslant}$ is $\geqslant$ and
$\overline{<}$ is $>$,
and $\Box \Phi \eqdef \false \Release \Phi$.

For an MC $\cM$ with states
labelled by $\cL \colon S \to \AP $ 
we use the standard semantics.
We only state the semantics of the
probability, expectation, and comparison operators here. 
For each state $s\in S$,
$ s \models_{\cM} \POp (\varphi) $ iff
$\Prob^{\cM}_s(\varphi) \bowtie c$,
and
$s \models_{\cM} \COp{\Prob}(\varphi_1,\bowtie, \varphi_2)$ iff
  $\Prob^{\cM}_s (\varphi_1) \bowtie \Prob^{\cM}_s (\varphi_2)$.
Here
$\Prob^{\cM}_s(\varphi)$ is short for
$\Pr^{\cM}_s\{\, \pi\in\Paths(s) : \pi\models_{\cM} \varphi \,\}$.
Furthermore,
$s \models_{\cM} \EOp (\rho)$ iff
$\Expect^{\cM}_s \bigl( \rho^{\cM} \bigr) \bowtie r$, and
$s \models_{\cM} \COp{\Expect}(\rho_1,\bowtie, \rho_2)$ iff
$\Expect^{\cM}_s \bigl(\rho_1^{\cM} \bigr) \bowtie 
 \Expect^{\cM}_s \bigl(\rho_2^{\cM} \bigr)$,
where $\Expect^{\cM}_s(\cdot)$ denotes the expected value 
of the respective random variable.
For detailed semantics of the expectation operators, see%
\citeAppendix{~Appendix \ref{app:semantics-expectation}. }%
\citeExtended{~\cite{GandALF-extended}. }%
We write $\cM \models \Phi$ iff $\sinit \models_{\cM} \Phi$.

\tudparagraph{1.0ex}{Notation: PCTL+EC and sublogics.}
We use PCTL to refer to unaugmented
probabilistic computation tree logic. If we
add only the expectation operator we write PCTL+E,
and, analogously, PCTL+C if we only add
the comparison operator for probabilities.
PCTL+EC denotes the full logic defined above.

\tudparagraph{1.0ex}{DAG-representation and length of formulas.}
We consider for any PCTL+EC state formula the
\emph{directed acyclic graph} (DAG) 
representing its syntactic structure.
Each node of the DAG represents
one of the sub-state formulas.
The use of a DAG rather than the syntax
tree allows the representation of subformulas that
occur several times in
the formula $\Phi$ by a single node.
The leaves of the DAG can be
the Boolean constant $\true$, and atomic propositions.
The inner nodes of the DAG, e.\,g., of a PCTL formula, are
labelled with one of the operators
$\wedge$, $\neg$, $\POp[\bowtie c](\,\cdot \Until \cdot\,)$,
$\POp[\bowtie c](\neXt \,\cdot\,)$.
Nodes labelled with $\neg$ and $\POp[\bowtie c](\neXt \,\cdot\,)$
have a single outgoing edge, while
nodes labelled with $\wedge$ or $\POp[\bowtie c](\,\cdot \Until \cdot\,)$
have two outgoing edges.
For the above-mentioned extensions of PCTL
the set of possible inner node labels is extended accordingly.
For example,  a node $v$ representing the PCTL+C formula
$\COp{\Pr}(\neXt \Phi_1, \bowtie, \Phi_2 \Until \Phi_3)$
has three outgoing edges. If $\Phi_1=\Phi_2$ then there are
two parallel edges from $v$ to a node representing $\Phi_1$.
The length of a PCTL+EC formula is defined as the number of nodes in its
DAG.

\makeatletter{}
\section{Fraction-free Gaussian elimination}
\label{sec:gauss}

Given a pMC $\fM$ as in Section \ref{sec:prelim},
the probabilities $\Prob^{\fM(\overline{x})}_s(\Diamond a)$
for reachability conditions are rational functions 
and computable via Gaussian elimination. 
As stated in the introduction,
this has been originally observed in \cite{Daws05,LanMagSchTroina07}
and realized, e.\,g., in the tools PARAM \cite{PARAM-HHWZ10} 
and Storm \cite{DJJCVBKA-CAV15,DJKV-CAV17}
together with techniques based on gcd-computations on multivariate
polynomials.
In this section, we discuss the potential of fraction-free
Gaussian elimination as an alternative,
which is well-known in mathematics
\cite{Bareiss72,GeCzLa93}, but to the best of our knowledge, 
has not yet been considered in the context of pMCs.

While the given definitions allow for rational functions
in the transition probability functions of (augmented) pMCs,
we will focus on polynomial (augmented) pMCs throughout
the remainder of the paper.
Generally, a linear equation systems containing rational functions as
coefficients can be rearranged to one containing only polynomials
by multiplying each line
with the common denominator of the respective rational functions.
Due to the multiplications this involves the risk of a blow-up
in the coefficient size.
To avoid this we
add variables in the following way.
Let $\fM = (S, \sinit, E, \bfP, \fC)$
be an (augmented) pMC.
For all $(s, t)\in E$
introduce a fresh variable $x_{s,t}$.
By definition $\bfP(s,t) = \frac{f_{s,t}}{g_{s,t}}$
for some $f_{s,t},g_{s,t}\in \Rational[\overline{x}]$.
Let $\bfP'(s,t) = f_{s,t}\cdot x_{s,t}$ if $(s,t)\in E$,
 $\bfP'(s,t) = 0$ if $(s,t)\notin E$,
$\fC' = \fC \cup \{ g_{s,t}\cdot x_{s,t} = 1 : (s,t)\in E \}$.
Then $\fM' = (S, \sinit, E, \bfP', \fC')$ is a polynomial augmented pMC.

\tudparagraph{1.0ex}{Linear equation systems with polynomial coefficients.}
Let $x_1,\ldots,x_k$ be parameters,
$\overline{x}=(x_1,\ldots,x_k)$.
We consider linear equation systems of the form $A\cdot p = b$,
where $A = (a_{i,j})_{i,j = 1,\ldots,n}$ 
is a non-singular $n\times n$-matrix
with $a_{i,j}=a_{i,j}(\overline{x}) \in \Rational[\overline{x}]$.
Likewise, $b = (b_i)_{i = 1,\ldots,n}$ is a vector of length $n$
with $b_i = b_i(\overline{x}) \in \Rational[\overline{x}]$.
The solution vector
$p = (p_i)_{i=1,\ldots,n}$ is a vector of rational functions
$p_i=f_i/g_i$ with $f_i,g_i \in \Rational[\overline{x}]$.
By Cramer's rule we obtain
$p_i = \frac{\det(A_i)}{\det(A)}$,
where $\det(A)$ is the determinant of $A$,
and $\det(A_i)$ the determinant of the matrix obtained when
substituting the $i$-th column of $A$ by $b$.
If the coefficients of $A$ and $b$ have at most degree $d$, the
Leibniz formula
implies that $f_i$ and $g_i$ have at most
degree $n\cdot d$. 

\begin{lemma}
\label{lemma:exp-many-monomials}
There is a family $(\fM_k)_{k\geqslant 2}$ 
of acyclic linear pMCs where
$\fM_k$ has $k$ parameters and $n=k{+}3$ states, including distinguished
states $s_0$ and $\goal$, such that $\Pr_{s_0}^{\fM(\overline{x})}(\Diamond \goal)$
is a polynomial for which even the shortest sum-of-monomial representation has 
$2^k$ monomials.
\end{lemma}

\begin{algorithm}[t]
\caption{One-step fraction-free Gaussian elimination \cite{Bareiss72}}
  \label{alg:gauss} 
  \label{algo:gauss}
\begin{algorithmic}[1]
  \Procedure{FractionFreeGauss}{$A = (a_{ij})_{i,j = 1,\ldots , n}$,
                                      $b = (b_i)_{i = 1,\ldots,n}$}
    \State $a_{0,0} = 1$ %
    \For{$m=1,\ldots,n {-} 1$} {\footnotesize\Comment triangulation, assuming $a_{m,m}\neq 0$}
      \For{$i =m {+} 1, \ldots , n$}
        \For{$j = m {+} 1, \ldots ,n$}
          \State\label{step:div1} $a_{i,j} = \bigl( a_{m,m} \cdot a_{i,j} - a_{i,m} \cdot a_{m,j}\bigr) / a_{m-1,m-1}$ {\footnotesize\Comment exploit exact divisibility by $a_{m-1,m-1}$}
        \EndFor
        \State\label{step:div2} $b_{i} = \bigl( a_{m,m} \cdot b_{i} - a_{i,m} \cdot b_{m}\bigr) / a_{m-1,m-1}$ {\footnotesize\Comment exploit exact divisibility by $a_{m-1,m-1}$}
        \State $a_{i,m} = 0$
      \EndFor
    \EndFor
    \For{$m = n{-}1, \ldots, 1$} {\footnotesize\Comment back substitution}
      \State\label{step:div3} $b_m = \bigl( a_{n,n} \cdot b_m - \sum_{i = m +1}^n a_{m,i} \cdot b_{i} \bigr) / a_{m,m}$  {\footnotesize\Comment exploit exact divisibility by $a_{m,m}$}
    \EndFor
    \State \textbf{return} $ \bigl( b_i / a_{n,n} \bigr) _{i = 1,\ldots, n}$ {\footnotesize\Comment rational solution functions}
  \EndProcedure
\end{algorithmic}
\end{algorithm}

\tudparagraph{1.0ex}{One-step fraction-free Gaussian elimination} 
is a variant of fraction-free Gaussian elimination
that allows for divisions which are known 
to be exact at the respective point in the algorithm.
When using \emph{na\"ive fraction-free Gaussian elimination}
the new coefficients after the $m$-th step, $m = 1,\ldots ,n-1$,
are computed as
$a^{(m)}_{i,j} = a^{(m-1)}_{i,j} a^{(m-1)}_{m,m} - a^{(m-1)}_{i,m} a^{(m-1)}_{m,j} $
for $i,j = m + 1,\ldots, n$, where $a^{(0)}_{i,j} = a_{i,j}$.
When applied to systems with polynomial coefficients this
results in the degree doubling after each step, so the
degree grows exponentially.
In a step of \emph{one-step fraction-free Gaussian elimination}
(see Algorithm~\ref{algo:gauss}),
the computation of the coefficients changes to
$a^{(m)}_{i,j}
  = \bigl(\, a^{(m-1)}_{i,j} a^{(m-1)}_{m,m} - a^{(m-1)}_{i,m} a^{(m-1)}_{m,j} \,\bigr)
    / a^{(m-1)}_{m-1,m-1}$
with $a^{(0)}_{0,0} = 1$.
Using Sylvester's identity one can prove that
$a^{(m)}_{i,j}$ is again a polynomial, and that
$a^{(m-1)}_{m-1,m-1}$ is in general the maximal
possible divisor.
The $b_i$ are updated analogously.
If the maximal degree of the
initial coefficients of $A$ and $b$
is $d$, this technique therefore guarantees, that
after $m$ steps the degree of the coefficients
is at most $(m{+}1)\cdot d$,
i.\,e., it grows linear in $d$ during the procedure.
For polynomials the division by $ a^{(m-1)}_{m-1,m-1} $ can be done using
standard polynomial division.
The time-complexity of the exact multivariate polynomial division
in this case is in each step
$\cO\bigl( \operatorname{poly}(md)^k \bigr)$, so for the full
one-step fraction-free Gaussian elimination it is
$\cO\bigl( \operatorname{poly}(n,d)^k \bigr)$.

Proposition 4.3 in \cite{LanMagSchTroina07} 
states that the rational functions $p_i=f_i/g_i$ 
for reachability probabilities in pMC 
with a representation of the polynomials
$f_i$, $g_i$ as sums of monomials 
(called normal form in \cite{LanMagSchTroina07})
are computable in polynomial time.
This contradicts Lemma \ref{lemma:exp-many-monomials}
which shows that the number of monomials
in the representation of a reachability
probability as a sum of monomials
can be exponential in the number of parameters.
However, the statement is correct for the univariate case.

\begin{lemma}
  \label{lemma:reach-prob-univariate}
  Let $\fM$ be a polynomial pMC over a single parameter and $T$ a set of states.
  Then, the rational functions for the reachability probabilities
  $\Pr^{\fM}_s(\Diamond T)$ are computable in polynomial time.
  Analogously, rational functions for 
    the expected accumulated 
    weight until reaching $T$ or the expected mean payoff
    are computable in polynomial time.
\end{lemma}

Note that the degrees of the polynomials $a_{i,j}^{(m)}$ and $b_j^{(m)}$
computed by one-step fraction-free Gaussian elimination
for reachability probabilities are bounded by $(m{+}1)\cdot d$,
where $d=\max_{s,t\in S} \degree(\bfP(s,t))$,
so the polynomials have representations as sums of at most $md{+}1$ monomials.
In particular, the degree and representation size of the final polynomials 
$f_s=b_s^{(n)}$ and $g_s=a_{s,s}^{(n)}$ for the
rational functions $\Pr^{\fM(x)}_s(\Diamond \goal)=f_s/g_s$ 
is in $\cO(n d)$ where $n$ is the number of states of $\fM$.
Another observation concerns the case where only the right-hand side of
the linear equation system is parametric.
Systems of this form occur, e.\,g., when considering
expectation properties for
MCs with parametric weights.
\begin{lemma}
  \label{lemma:right-hand-param}
  Let $A \cdot p = b$ be a parametric linear equation system as defined above
  where $A$ is parameter-free. Then the solution vector
  $p=(p_i)_{i=1,\ldots,n}$ consist of
  polynomials of the form 
  $p_i = \sum_{i = 1}^n \beta_i \cdot b_i$ with
  $\beta_i\in \Rational$ and can be computed in
  polynomial time. 
\end{lemma}

\tudparagraph{1.0ex}{Stratification via SCC-decomposition.}
It is well known~(e.\,g., \cite{CBGK08,JansenCVWAKB14}) that for
probabilistic/parametric model checking a decomposition into strongly-connected
components (SCCs) can yield significant performance benefits due to
the structure of the underlying models.
We have adapted the one-step fraction-free Gaussian elimination
approach by a preprocessing step that permutes the matrix according to
the topological ordering of the SCCs. 
This results in the coefficient matrix already having a stair-like form
at the start of the algorithm.
In the triangulation part of the algorithm, each SCC 
can now be considered separately,
as non-zero entries below the main diagonal only occur within each SCC.
While the back-substitution in the general one-step fraction-free elimination
will result in each entry on the main diagonal being equal to the last,
this property is now only maintained within the SCCs.
Formally, this means that the back substitution step 
in Algorithm \ref{alg:gauss} is replaced by the following:
\begin{eqnarray*}
    b_m &= &
     \Bigl(a^*\!(\text{current SCC}) \cdot b_m -
           \sum\limits_{i = m + 1}^n a_{m,i} \cdot b_i \cdot 
           \dfrac{a^*\!(\text{current SCC})}
          {a^*\!(\text{SCC at $i$})}\Bigr)
          \; / \; a_{m,m}
\end{eqnarray*}
where $a^*\!(\text{SCC at $n$}) = a_{n,n}$, and, for $i = 1,\ldots, n{-}1$,
   $a^*\!(\text{SCC at $i$}) = 
     a^*\!(\text{SCC at $i {+} 1$})$
     if the $i$-th and $(i{+}1)$-st state belong to the same SCC
and  
   $a^*\!(\text{SCC at $i$}) = 
    a_{i,i} \cdot a^*\!(\text{SCC at $i {+} 1$})$
    otherwise.
Intuitively, $a^*\!(\text{SCC at $i$})$ is the product of the
$a$'s on the diagonal corresponding to the last states in the current
SCC and the SCCs below. 
Of course, the return statement also has to be adjusted accordingly.
The advantage of this approach is that the polynomials in the 
rational functions
aside from the ones in the first strongly connected component will have
an even lower degree. 
 
\makeatletter{}\tudparagraph{1.0ex}{Implementation and experiments.}
For a first experimental evaluation of the one-step fraction-free Gaussian
elimination approach (\geff) in the context of probabilistic model checking,
we have implemented this method (including the SCC decomposition and
topological ordering described above) as an alternative solver for 
parametric linear equation systems in the state-of-the-art
probabilistic model checker Storm~\cite{DJKV-CAV17}.
We compare \geff{} against the two solvers provided by Storm
(v1.0.1)
for solving parametric equation systems,
i.\,e., the solver based on the \eigen{} linear algebra
library\footnote{\url{http://eigen.tuxfamily.org/}}
and
on state elimination (\selim{})~\cite{HHZ-STTT11}.
Both of Storm's solvers use
partially factorized representations of the rational functions
provided by the CArL library%
\footnote{\url{https://github.com/smtrat/carl}}.
This approach, together with caching, was shown~\cite{JansenCVWAKB14} to be beneficial
due to improved performance of the gcd-computations during the 
simplification steps.

It should be noted that our implementation is intended to provide
first results that allow to gauge whether the fraction-free method, by
avoiding gcd-computations, can
be beneficial in practice and is thus rather na\"ive in certain aspects.
As an example, it currently relies on a dense matrix representation,
with performance improvements for larger models to be expected from switching to
sparse representations as used in Storm's \eigen{} and \selim{} solvers.
In addition to the fraction-free approach, our solver can also
be instantiated to perform a straight-forward Gaussian elimination,
using any of the representations for rational functions provided by
the CArL library.
In all our experiments, we have compared the solutions obtained by the
different solvers and verified that they are the same.

\tudparagraph{1.0ex}{Experimental studies.}
The source code of our extension of Storm and the artifacts
of the experiments are available online.%
\footnote{\url{http://wwwtcs.inf.tu-dresden.de/ALGI/PUB/GandALF17/}}
As our \geff{} implementation is embedded as an alternative solver in
Storm, we mainly report the time actually spent for solving the
parametric equation system, as the other parts of model checking
(model building, precomputations) are independent of the chosen solver.
For benchmarking, we used a machine with
two Intel Xeon E5-2680 8-core CPUs at 2.70GHz and with 384GB RAM,
a time out of 30 minutes and a memory limit of 30GB.
All the considered solvers run single-threaded.
We have considered three different classes of case studies for
experiments.

\tudparagraph{1.0ex}{Complete pMC.}
As a first experiment to gauge the efficiency in the presence of a
high ratio of parameters to states, we considered a family of pMCs
with a complete graph structure (over $n$ states) and one parameter
per transition, resulting in $n\cdot(n+1)$ parameters 
(for details see%
\citeAppendix{~Appendix~\ref{app:experiments}). }%
\citeExtended{~\cite{GandALF-extended}). }%

\begin{table}
\caption{Statistics for ``complete pMC''.
Matrix rows and number of distinct parameters, as well as time for
solving the parametric equation system per solver. For $n=7$, all solvers timed out (30min).}
\centering
\begin{tabular}{rrr|rrr|r}
$n$  & rows & param. & \eigen{} & \selim{} &
\geff{} & \geffred{} \\\hline
4 & 4 & 20 
 & \psec{0.469}  %
 & \psec{0.643}  %
 & \psec{0.055}  %
 & \psec{0.013}  %
 \\
5 & 5 & 30 
 & \psec{44.470}  %
 & \psec{42.088}  %
 & \psec{2.131}  %
 & \psec{1.523}  %
 \\
6 & 6 & 42  
 & \timeout    %
 & \timeout    %
 & \psec{221.271}  %
 & \psec{21.528}  %
 \\\hline
\end{tabular}
\label{tab:complete-pdtmc}
\end{table}

Table~\ref{tab:complete-pdtmc} depicts statistics for the
corresponding computations, for the two standard solvers in Storm
(\eigen{} and \selim{}), as well as our fraction-free
implementation (\geff{}). For \selim{}, we always use
the default elimination order (forward). The time for \geff{} corresponds
to the time until a solution rational function (for all states) is obtained.
As the numerator and denominator of these rational
functions are not necessarily coprime, for comparison purposes we
list as well the time needed for simplification (\geffred{}) 
via division by the gcd.
As can be seen, here, the fraction-free approach
significantly outperforms Storm's standard solvers and scales to a
higher number of parameters. We confirmed using profiling that
the standard solvers indeed spend most of the time in
gcd-computations.

\tudparagraph{1.0ex}{Multi-parameter Israeli-Jalfon self-stabilizing.}
The benchmarks used to evaluate parametric model checking
implementations in previous papers tend to be scalable in the number
of components but use a fixed number of parameters, usually 2. To
allow further experiments with an increasing number of parameters, we
considered a pMC-variant of the Israeli-Jalfon self-stabilizing protocol
with $n$ processes, $k$ initial tokens and $n$ parameters (for details
see%
\citeAppendix{~Appendix~\ref{app:experiments}). }%
\citeExtended{~\cite{GandALF-extended}). }%

\begin{table}
\caption{Statistics for ``Israeli-Jalfon'', with strong bisimulation quotienting.
Matrix rows and number of distinct parameters, as well as time for
solving the parametric equation system per solver.}
\centering
\begin{tabular}{rrrr|rrrr|r}
$n$  & $k$ & rows & param. & \eigen{} & \selim{} &
\gefac{} & 
\geff{} & \geffred{} \\\hline
4 & 3 & 21 & 4
 & \psec{1.014}  %
 & \psec{0.864}  %
 & \psec{0.730}  %
 & \psec{0.159}  %
 & \psec{0.204}  %
 \\
4 & 4 & 15 & 4
 & \psec{0.938}  %
 & \psec{0.575}  %
 & \psec{0.577}  %
 & \psec{0.156}  %
 & \psec{0.128}  %
 \\
5 & 2 & 16 &  5
 & \psec{19.127}  %
 & \psec{30.831}  %
 & \psec{29.459}  %
 & \psec{9.364}  %
 & \psec{0.331}  %
 \\
5 & 3 & 36 &  5
 & \psec{360.425}  %
 & \psec{747.320}  %
 & \psec{172.156}  %
 & \psec{485.775}  %
 & \psec{95.923}  %
 \\
5 & 4 & 51 &  5 
 & \psec{457.548}  %
 & \psec{1542.973}  %
 & \psec{442.797}  %
 & \psec{614.013}  %
 & \psec{742.694}  %
 \\
5 & 5 & 31 &  5
 & \psec{368.700}  %
 & \psec{1597.291}  %
 & \psec{252.915}  %
 & \psec{621.997}  %
 & \psec{414.717}  %
 \\\hline
\end{tabular}
\label{tab:ij}
\end{table}

Table~\ref{tab:ij} depicts the time spent for computing the
rational functions for several instances. As can
be seen, the fraction-free approach is competitive for the smaller
instances, with performance between the \eigen{} and
\selim{} solvers for the larger instances. We have also
included running times for \gefac{}, i.\,e., for our
na\"ive implementation of Gaussian elimination using the
representation for rational functions as used by Storm for the
standard solvers, including automatic gcd-based simplification after
each step to ensure that numerator and denominator are coprime.
\gefac{} operates
on the same, topologically sorted matrix as the fraction-free
\geff{}. Curiously, \gefac{} is able to outperform the
\eigen{} solver for some of the larger instances. We believe this
is mainly due to differences in the matrix permutation and their
effect on the elimination order, which is known to have a large impact
on performance (e.\,g.,~\cite{DJJCVBKA-CAV15}).

\tudparagraph{1.0ex}{Benchmark case studies from \cite{DJJCVBKA-CAV15}.}
Furthermore, we considered several case study instances that were used
in~\cite{DJJCVBKA-CAV15} to benchmark parametric model checkers,
namely the \emph{brp}, \emph{crowds}, \emph{egl}, \emph{nand},
\emph{zeroconf} models.
Table~\ref{tab:bench-selected} depicts statistics for selected
instances, for further details see%
\citeAppendix{~Appendix~\ref{app:experiments}. }%
\citeExtended{~\cite{GandALF-extended}. }%
The application of bisimulation quotienting often has a large impact
on the size of the linear equation system, so we
performed experiments with and without quotienting.
For \emph{crowds}, bisimulation quotienting was particularly effective, with all
considered instances having a very small state space and negligible
solving times. For the non-quotiented instances, Storm's standard
solvers outperform \geff{}. For the \emph{zeroconf} instance in
Table~\ref{tab:bench-selected}, \geff{} is competitive.
Note that the models in the
\emph{brp}, \emph{egl} and \emph{nand} case studies are acyclic and
that the parametric transition probabilities and rewards are
polynomial. As a consequence, the gcd-computations used in Storm's
solvers don't impose a significant overhead as the rational functions
during the computation all have denominator polynomials of degree zero.

\begin{table}
\caption{Selected statistics for the benchmarks of \cite{DJJCVBKA-CAV15}.
Matrix rows and number of distinct parameters, as well as time for
solving the parametric equation system per solver.}
\label{tab:bench-selected}
\centering
\begin{tabular}{rrr|rrr|r}
model  & rows & param. & \eigen{} & \selim{} &
\geff{} & \geffred{} \\\hline
Crowds (3,5), weak-bisim%
 & 40  %
 & 2   %
 & \psec{0.077}  %
 & \psec{0.062}  %
 & \psec{0.018}  %
 & \psec{0.126}  %
\\
Crowds (5,5), weak-bisim%
 & 40  %
 & 2   %
 & \psec{0.076}  %
 & \psec{0.059}  %
 & \psec{0.018}  %
 & \psec{0.109}  %
\\ 
Crowds (10,5), weak-bisim %
 & 40  %
 & 2   %
 & \psec{0.077}  %
 & \psec{0.060}  %
 & \psec{0.019}  %
 & \psec{0.109}  %
\\
Crowds (3,5)  %
 & 715  %
 & 2   %
 & \psec{0.989}  %
 & \psec{0.800}  %
 & \psec{11.435}  %
 & \psec{63.386}  %
\\
Crowds (5,5) %
 & 2928  %
 & 2   %
 & \psec{6.357}  %
 & \psec{5.505}  %
 & \psec{1271.954}  %
 & \timeout
\\
Crowds (10,5) %
 & 25103  %
 & 2   %
 & \psec{139.818}  %
 & \psec{173.147}  %
 & \timeout  %
 & ---
\\
\hline
Zeroconf (1000) %
 & 1002  %
 & 2   %
 & \psec{81.027}  %
 & \psec{45.007}  %
 & \psec{49.427}  %
 & \psec{11.352}  %
\\
\hline
\end{tabular}
\end{table}

Overall, the experiments have shown that there are instances where the
fraction-free approach can indeed have a positive impact on
performance. Keeping in mind that our implementation has not yet been
significantly optimized, we believe that the fraction-free approach is
an interesting addition to the gcd-based solver approaches. In particular,
the application of better heuristics for the order of processing (i.\,e., the
permutation of the matrix) could still lead to significant performance
increases.

\makeatletter{}
\section{Complexity of the PCTL+EC model checking problem}
\label{sec:theory}

We now study the complexity of the following variants of the PCTL+EC 
model checking problem. 
Given an augmented pMC $\fM = ( S , \sinit, E, \bfP, \fC)$ and a
PCTL+EC (state) formula $\Phi$:
\begin{center}
  \begin{tabular}{ll}
    (All) &
    Compute a representation of the set of all satisfying
    parameter valuations, 
    \\

    & 
    i.\,e., the set of all admissible parameter
    valuations $\overline{\xi}\in X$ such that
    $\fM(\overline{\xi}) \models \Phi$.
    \\[1ex]
    (MC-E) &
    Does there exist a valuation $\overline{\xi}\in X$
		such that $\fM(\overline{\xi}) \models \Phi$?
    \\[1ex]
    (MC-U) &
    Does $\fM(\overline{\xi}) \models \Phi$
		hold for all admissible valuations $\overline{\xi}\in X$?
  \end{tabular}
\end{center}
(MC-E) and (MC-U) are essentially duals of each other.
Note that the answer for the 
universal variant (MC-U) is obtained by the negation of the answer for
(MC-E) with formula $\neg\Phi$, and vice versa.
In what follows, we shall concentrate on (All) and the existential
model checking problem (MC-E).

\tudparagraph{1.0ex}{Computing all satisfying parameter valuations.}
As before, 
$X=X_{\fM}$ denotes the set of admissible valuations.
In what follows, let $\chi$ be the conjunction
of the polynomial constraints in $\fC$ as well as the constraints
$\sum_{t\in S} \bfP(s,t)=1$ for each non-trap state $s\in S$, and
$0 < \bfP(s,t)$ for each edge $(s,t)\in E$.
We then have $\overline{\xi}\models \chi$ if and only if 
$\overline{\xi}$ is admissible, i.\,e., $\overline{\xi}\in X$.

Let $\Phi$ be a PCTL+EC formula. 
The \emph{satisfaction function}
$\Sat_{\fM}(\Phi) \colon X \to 2^{S}$ is defined by:
\begin{align*}
	\Sat_{\fM}(\Phi)(\overline{\xi}) 
    & \stackrel{\text{\tiny def}}{=}
        \bigl\{ \,
		      s\in S : s\models_{\fM(\overline{\xi})} \Phi
        \, \bigr\}
	  = \Sat_{\fM(\overline{\xi})} (\Phi) 
\end{align*}
We now present an algorithm to compute a symbolic representation 
of the satisfaction function that groups valuations
with the same corresponding satisfaction set together.
More precisely, we deal with a representation
of the satisfaction function $\Sat_{\fM}(\Phi)$
by a finite set $\tSat_{\fM}(\Phi)$ 
of pairs $(\gamma,T)$ where $\gamma$ is a Boolean combination
of constraints and $T \subseteq S$ such that
(i) $(\gamma,T)\in \tSat_{\fM}(\Phi)$ and $\overline{\xi}\models \gamma$ 
implies
$T = \Sat_{\fM}(\Phi)(\overline{\xi})$,
and
(ii) whenever $T = \Sat_{\fM}(\Phi)(\overline{\xi})$ then
there is a pair $(\gamma,T)\in \tSat_{\fM}(\Phi)$ such that
$\overline{\xi}\models \gamma$.

Given the DAG representation of the PCTL formula $\Phi$,
we follow the standard model checking procedure for CTL-like
branching-time logics and compute $\tSat_{\fM}(\Psi)$ for the
subformulas $\Psi$ 
assigned to the nodes in the DAG for $\Phi$ in a bottom-up manner.
As the leaves of the DAG can be atomic propositions $a$ or the
formula $\true$, the base cases are
$\tSat_{\fM}(\true) = \bigl\{ (\chi,S) \bigr\}$, and
\mbox{$\tSat_{\fM}(a) =  \bigl\{ \, 
        (\chi,\{s\in S : a \in \cL(s)\}) \, \bigr\}$}.
Consider now the inner node $v$ of the DAG for $\Phi$
labelled by the outermost operator of the subformula $\Psi$.
Suppose that the children of $v$ have already been treated,
so when computing
$\tSat_{\fM}(\Psi)$ the satisfaction
sets of the proper subformulas of $\Psi$ are known.
If $v$ is labelled by $\neg$ or $\wedge$, i.\,e.,
$\Psi = \neg\Psi'$ or $\Psi = \Psi_1 \wedge \Psi_2$, then
$\tSat_{\fM}(\Psi) = 
  \bigl\{ \, (\gamma,S \setminus T) :
    (\gamma,T)\in \tSat_{\fM}(\Psi')\, \bigr\}$
respectively
$\tSat_{\fM}(\Psi) =
  \bigl\{ \ (\gamma_1 \wedge \gamma_2,T_1 \cap T_2) \ : \ 
    (\gamma_i,T_i)\in \tSat_{\fM}(\Psi_i),\  i=1,2 \ \bigr\}$.
If $\Psi = \PrOp_{\bowtie c}(\Psi_1 \Until \Psi_2)$, then
\begin{align*}
  \tSat_{\fM}(\Psi)
    & \ = \
      \bigl\{ \
            (\gamma_1 \wedge \gamma_2 \wedge \delta_{\gamma_1,T_1,\gamma_2,T_2,R}) \ 
            : \
            (\gamma_1,T_1)\in \tSat_{\fM}(\Psi_1), \ 
            (\gamma_2,T_2)\in \tSat_{\fM}(\Psi_2), \ 
            R \subseteq S 
      \ \bigr\}
\end{align*}
where
$\delta_{\gamma_1,T_1,\gamma_2,T_2,R}$ 
is the conjunction of the constraints
  $\Pr^{\fM}_{s}(T_1 \Until T_2) \bowtie c$
for each state $s\in R$, and 
  \mbox{$\Pr^{\fM}_{s}(T_1 \Until T_2) \;\not\bowtie c$}
for each state $s\in S \setminus R$.
Here, $\Pr^{\fM}_{s}(T_1 \Until T_2)$ is the rational function
that has been computed using (i) a graph analysis to determine the set $U$
of states $s$ with $s \models \exists (T_1 \Until T_2)$ and
(ii) fraction-free Gaussian elimination (Section \ref{sec:gauss}) 
to compute the rational functions $\Pr_s^{\fN}(\Diamond T_2)$ 
in the pMC $\fN$ resulting from
$\fM$ by turning the states in $(S \setminus U) \cup T_2$
into traps.
If $f_s$ and $g_s$ are polynomials computed by
fraction-free Gaussian elimination 
such that 
$\Pr^{\fM}_{s}(T_1 \Until T_2) = f_s/g_s$ then
$\Pr^{\fM}_{s}(T_1 \Until T_2) \bowtie c$ 
is a shortform notation for $f_s - c \cdot g_s \bowtie 0$.
The treatment of 
$\PrOp_{\bowtie c}(\neXt \Psi)$ and the expectation
operators is similar, and can be found in%
\citeAppendix{~Appendix~\ref{sec:All-SAT}. }%
\citeExtended{~\cite{GandALF-extended}. }%
After treating a node of the DAG,
we can simplify the set $\tSat_{\fM}(\Psi)$
by first removing all pairs $(\gamma,T)$ where
$\gamma$ is not satisfiable (using algorithms
for the existential theory of the reals),
and afterwards combining all pairs with the same $T$-component,
that is, instead of $m$ pairs 
$(\gamma_1,T),\ldots,(\gamma_m,T)\in \tSat_{\fM}(\Psi)$,
we consider a single pair $(\gamma_1 \vee \ldots \vee \gamma_m,T)$.
To answer question (All), the algorithm finally returns the 
disjunction of all formulas $\gamma$
with $\sinit \in T$ for $(\gamma,T)\in\tSat_{\fM}(\Phi)$.

\tudparagraph{1.0ex}{Complexity bounds of (All) and (MC-E).}
The existential theory
of the reals is known to be in PSPACE and NP-hard, and there is an upper
bound on the time-complexity, namely $\ell^{k+1}\cdot d^{\cO(k)}$ where
$\ell$ is the number of constraints, $d$ the maximum degree
of the polynomials in the constraints,
and $k$ the number of parameters \cite{BaPoRo08}.
Recall from Section \ref{sec:gauss} that a known upper bound on the
time-complexity of one-step fraction-free Gaussian elimination
is $\cO\bigl( \operatorname{poly}(n,d)^k \bigr)$, where
$n$ is the number of equations, $d$ the maximum degree of the initial
coefficient polynomials, and $k$ the number of parameters.
Combining both approaches, the one-step fraction-free Gaussian elimination
for solving linear equation systems with polynomial coefficients,
and the existential theory of the reals for treating satisfiability
of conjunctions of polynomial constraints, one directly obtains the following
bound for the computational complexity of 
PCTL+EC model checking on augmented polynomial pMCs.
Note that this assumes that the number of constraints in $\fC$
is at most polynomial in the size of $S$.

\begin{theorem}[Exponential-time upper bound for problem (All)]
 \label{thm:complexity-PCTL-multivariate}
  Let $\Phi$ be a PCTL+EC formula.
  Given an augmented polynomial pMC $\fM$, where
  the maximum degree of transition probabilities 
  $\bfP(s,t)$,
  and polynomials in the constraints
   in $\fC$ is $d$, 
  a symbolic representation of
  the satisfaction function $\Sat_{\fM}(\Phi)$ is computable
  in time 
  $\cO\bigl( |\Phi|\cdot 
    \operatorname{poly}\bigl(
      \text{size}(\fM), d
    \bigr)^{k\cdot |\Phi|_{\POp[],\EOp[],\CompOp}} \bigr)$,
  where $|\Phi|_{\POp[],\EOp[],\CompOp}$ is the
  number of probability, expectation and comparison operators in $\Phi$.
\end{theorem}

\begin{theorem}[PSPACE upper bound for problem (MC-E)]
 \label{PSPACE-multivariate}
   The existential PCTL+EC model checking problem (MC-E)
   for augmented pMC
   is in PSPACE.
\end{theorem}

\begin{sketch}
  The main idea of a polynomially space-bounded
  algorithm is to guess nondeterministically sets $T_{\Psi}$ of states 
  for the subformulas $\Psi$ where the outermost operator is a
  probability, expectation or comparison
  operator, and then apply a polynomially space-bounded  
  algorithm for the existential theory of the reals
  \cite{BaPoRo08}
  to check whether there is a parameter valuation $\overline{\xi}$
  such that $T_{\Psi}=\Sat_{\fM}(\Psi)(\overline{\xi})$ for all
  $\Psi$.
\end{sketch}

NP- and coNP-hardness of (MC-E) follow from results 
for IMCs \cite{SeViAg06,ChatSenHen08}.
More precisely, \cite{ChatSenHen08} provides a
polynomial reduction from SAT 
to the (existential and universal) PCTL model checking problem
for IMCs. 
In fact, the reduction of \cite{ChatSenHen08} does not require full PCTL,
instead Boolean combinations of simple probabilistic constraints
 $\PrOp_{\geqslant c_i}(\neXt a_i)$
without nesting  of the probability operators
are sufficient.
The following theorem strengthens this result
by stating NP-hardness of (MC-E) even for formulas 
$\POp[>c](\Diamond a)$ consisting of a single
probability constraint for a reachability condition.

\begin{theorem}[NP-hardness for single probabilistic operator, 
            multivariate case]
          \label{thm:NP-hard-multi}
	Given an augmented polynomial pMC\, $\fM$ 
        on parameters $\overline{x}$ with initial state
        $\sinit$ and an atomic
        proposition $a$, and a probability threshold 
        $c\in \Rational \,\cap\, ]0,1[$, 
        the problem to decide whether there exists
  $\overline{\xi} \in X$
	such that $\Pr^{\fM(\overline{\xi})}_{\sinit}(\Diamond a) > c$
        is NP-hard,
  even for acyclic pMCs 
    with the assigned transition probabilities being either constant,
    or linear in one parameter, i.\,e.,
    $\bfP (s,t) \in \bigcup_{i = 1}^k \Rational [x_i]$,
    $\deg(\bfP(s,t))\leq 1$,
    for all $(s,t)\in E$, and
  where the polynomial constraints
  for the parameters $x_1,\ldots,x_k$ are
  of the form $f(x_i) \geqslant 0$ 
  with $f\in \Rational [x_i]$, $\deg(f) \leq 2$.
\end{theorem}
 
\makeatletter{}\tudparagraph{1.0ex}{Univariate pMCs.}
In many scenarios, 
the number of variables has a fixed bound
instead of increasing with
the model size.
We consider here the case of \emph{univariate} pMC, i.\,e.,
pMC with a single parameter.

\begin{theorem}[PCTL+EC model checking without nesting in P, univariate case]
  \label{thm:upMC-nonnest}
	Let $\Phi$ be a PCTL+EC formula
  without nested probability, expectation or comparison operators,
	and let $\fM$ be a polynomial pMC on the single parameter $x$.
  The problem to decide whether there exists an
  admissible parameter valuation $\xi \in X$
  such that $\fM(\xi) \models \Phi$ is in P.
\end{theorem}

\begin{sketch}
If we restrict PCTL+EC to Boolean combinations of
probability, expectation, and comparison operators,
(MC-E) can be dealt with by first
computing polynomial constraints for $\sinit$ for
each probability, expectation, and comparison operator independently
(this can be done in polynomial time 
by Lemma \ref{lemma:reach-prob-univariate}),
and afterwards applying a polynomial-time algorithm for the
univariate existential theory of the reals \cite{BeKoRe86}
once to the appropriate Boolean combination of the
constraints.
\end{sketch}

\begin{theorem}[NP-completeness for full PCTL+EC, univariate case]
  \label{thm:np-complete-uni}
	Let $\Phi$ be a PCTL+EC formula, and let $\fM$ be a
	polynomial pMC on the single parameter $x$.
	The PCTL+EC model checking problem to decide whether there exists an
	admissible parameter valuation $\xi \in X$
	such that $\fM(\xi) \models \Phi$ is NP-complete.
  NP-hardness even holds for acyclic polynomial pMCs 
  and the
  fragment of PCTL+C that uses the
  comparison operator $\COp{\Pr}$, 
  but not the probability operator $\PrOp$,
  as well as for (cyclic) polynomial pMC in combination with PCTL.
\end{theorem}

\makeatletter{}
\tudparagraph{1.0ex}{(MC-E) for monotonic PCTL on univariate pMCs.}
The parameters in pMC typically have a fixed meaning, e.\,g., 
probability for the
occurrence of an error, in which case the probability to reach a 
state where an error has occurred is increasing in $x$.
This motivates the consideration of univariate  pMCs and 
PCTL formulas that are monotonic in the following sense.

Given a univariate polynomial pMC $\fM=(S,\sinit,E,\bfP)$,
let $E_+$ denote the set of edges $(s,t)\in E$ such that
the polynomial $\bfP(s,t)$ is monotonically increasing in $X$,
i.\,e., whenever $\xi_1,\xi_2\in X$ and
 $\xi_1 < \xi_2$ then $\bfP(s,t)(\xi_1)\leqslant \bfP(s,t)(\xi_2)$.
Let $S_+$ denote the set of states $s$ such that for each finite
path $\pi = s_0\, s_1 \ldots s_m$ with $s_m=s$ we have
$(s_i,s_{i+1})\in E_+$ for $i=0,1,\ldots,m{-}1$.

As $(s,t)\in E_+$ iff
there is no value $\xi \in \Real$ such that
$\xi \models \chi \wedge (\bfP(s,t)'<0)$,
the set $E_+$ is computable in polynomial time using
a polynomial-time algorithm for the univariate theory of the reals 
\cite{BeKoRe86}.
Here, $\chi$ is as before the
  Boolean combination of polynomial constraints characterizing the set 
  $X$ of admissible parameter values, and
  $\bfP(s,t)'$ is the first derivative of the polynomial $\bfP(s,t)$.
Thus, the set $S_+$ is computable in polynomial time.

\begin{lemma}
Let $\fM $ be a univariate polynomial pMC
and $\Psi$ a monotonic PCTL formula,
that is, $\Psi$ is in the PCTL fragment obtained by the following grammar:
\vspace*{-0.5em}
\begin{eqnarray*}
  \Phi    & \ ::= \ &  
   a \in S_+  \ \mid \ \Phi\wedge\Phi \
               \mid \ \Phi\vee\Phi \ \mid \ \POp[\geqslant c](\varphi) 
   \ \mid \ \POp[> c](\varphi) 
  \\
  \varphi &::= &
  \neXt\Phi \ \mid \ \Phi \Until \Phi \ \mid \ \Phi \Release \Phi
  \ \mid \ \ \Diamond \Phi \ \mid \  \ \Box \Phi
\end{eqnarray*}
where $c \in \Rational_{>0}$.
Then, $\Sat_{\fM(\xi_1)}(\Psi) \subseteq \Sat_{\fM(\xi_2)}(\Psi)$
for any two valuations $\xi_1$ and $\xi_2$ of $x$ with $\xi_1 < \xi_2$.
\end{lemma}

Hence, if $\Psi$ is monotonic then the satisfaction function
$X \to 2^S$, $\xi \mapsto \Sat_{\fM}(\Psi)(\xi) = \Sat_{\fM(\xi)}(\Psi)$
is monotonic.
For each monotonic PCTL formula $\Psi$
there exist $S_{\Psi} \subseteq S$ and $\xi_{\Psi} \in X$ such that
$\Sat_{\fM(\xi)}(\Psi) = S_{\Psi}$ for all $\xi \geqslant \xi_{\Psi}$
and $\Sat_{\fM(\xi')}(\Psi) \subseteq S_{\Psi}$ for all $\xi' < \xi_{\Psi}$.
To decide (MC-E) for a given monotonic formula $\Phi$, 
it suffices to determine the sets $S_{\Psi}$ 
for the sub-state formulas $\Psi$ of $\Phi$.
This can be done in polynomial time. Using this observation, we obtain:

\begin{theorem}[(MC-E) for monotonic PCTL on univariate pMC]
\label{thm:mon-uni}
Let $\fM = (S, \sinit, E, \bfP, \fC)$ be a univariate polynomial pMC on $x$,
and $\Phi$ a monotonic PCTL formula.
Then the model checking problem to decide whether there exists an
admissible parameter valuation $\xi$ for $x$ such that $\fM(\xi)\models \Phi$
is in P.
\end{theorem} 
\makeatletter{}\tudparagraph{1.0ex}{Model checking PCTL+EC on MCs with parametric weights.}
We now consider the case where $\cM$ is an ordinary Markov chain
augmented with a parametric weight function
$\wgt \colon S \to \Rational [\overline{x}]$.
Given a set $T \subseteq S$ such that $\Pr^{\cM}_s(\Diamond T)=1$
for all states $s \in S$,
the vector of the expected accumulated weights
$e = (\Expect_s^{\cM}(\accdiaplus T))_{s\in S}$
is computable
as the unique solution of a linear equation
system of the form $A \cdot e = b$,
where the matrix $A$ is non-parametric, and only the vector $b$
depends on $\overline{x}$.
By Lemma \ref{lemma:right-hand-param},
$\Expect_s^{\cM}(\accdiaplus T)$ is a polynomial of the form 
$\sum _{t\in S} \beta_{s,t}\cdot \wgt(t)$
with $\beta_{s,t}\in \Rational$ for all $s\in S$,
and can be computed in polynomial time.
The expected mean payoff for a given set $T$ is given by
$\Expect_s^{\cM}(\MeanPayoff(T)) =
  \sum_{\text{BSCC $B$}} \Pr_s^{\cM}(\Diamond B)\cdot
  \MeanPayoff(B)(T)$
where $\MeanPayoff(B)(T) = \sum_{t\in T} \zeta_t \cdot \wgt_T(t)$
with $\zeta_t$ being the steady-state probability for state $t$ inside $B$
(viewed as a strongly connected Markov chain), and
$\wgt_T(t)=0$ if $t \notin T$, $\wgt_T(t)=\wgt(t)$ for $t\in T$.
As the transition probabilities are non-parametric, the steady-state
probabilities are obtained as the unique solution of a 
non-parametric linear equation system.
So both types of expectations can be computed in polynomial time.
Unfortunately, the treatment of formulas with nested expectation operators
is more involved. 
Using the standard computation scheme that processes
the DAG-representation of the given PCTL+EC formula in a bottom-up manner
to treat inner subformulas first,
the combination of polynomial constraints after
the consideration of an inner node 
is still as problematic as in the pMC-case. Using
known algorithms for the existential theory of the reals
yields the following bound.

\begin{theorem}[Time complexity of 
   PCTL+EC model checking with parametric weights]
\label{PCTL+EC-weights}
 Let $\cM$ be an MC with parametric weights over $k$ parameters,
 and $\Phi$ a PCTL+EC formula.
 The problem (MC-E)
 is solvable in time
	$\cO\bigl(\, |\Phi|\cdot
		\operatorname{poly}\bigl(\text{size}(\cM),d\bigr)^{k
		\cdot |\Phi|_{\EOp[], \COp{\Expect}}} \,\bigr)$,
	where $|\Phi|_{\EOp[], \COp{\Expect}}$ is the number of
	expectation and expectation comparison operators in the formula,
	and $d$ the maximum degree of the polynomials assigned as weights.
\end{theorem}

If there is only one parameter,
the model checking for MCs with parametric weights is solvable
in polynomial time for the fragment of PCTL+EC without nested formulas
(cf.~Theorem \ref{thm:upMC-nonnest}).

\makeatletter{}
\section{Conclusion}
\label{sec:conc}

In this paper we revisited the model checking problem for pMC and PCTL-like
formulas. 
The purpose of the first part is to draw attention
to the fraction-free Gaussian elimination for computing rational functions for 
reachability probabilities, expected accumulated weights and expected
mean payoffs as an alternative to the gcd-based algorithms 
that have been considered
before and are known to suffer from the high complexity of gcd-computations for
multivariate polynomials.
The experiments with our (not yet optimized) implementation
indicate that such an approach can indeed be feasible and beneficial
in practice. We thus intend to refine this implementation in future
work, including research into further structural heuristics and the
potential of a combination with gcd-based simplifications at opportune
moments.

In the second part of the paper we studied the complexity of the
model checking problem for pMC and PCTL and its extension PCTL+EC
by expectation and comparison operators. 
We identified instances where the
model checking problem is NP-hard as well as
fragments of PCTL+EC where the model checking
problem is solvable in polynomial time.
The latter includes the model checking problem for
Boolean combinations of probability or
expectation conditions for univariate pMCs.
This result has been obtained using the fraction-free Gaussian
elimination to compute rational functions for reachability probabilities
or expected accumulated weights or expected mean payoffs, 
and polynomial time algorithms for the
theory of the reals over a fixed number of variables.
As the time complexity of 
the fraction-free Gaussian elimination is also polynomial
for matrices and vectors with a fixed number of parameters
and the polynomial-time decidability for the theory of the reals
also holds when the number of variables is fixed \cite{BeKoRe86},
Theorem \ref{thm:upMC-nonnest} also holds for pMC with a fixed
number of parameters.

\bibliographystyle{eptcs}
\bibliography{lit}

\end{document}